\documentclass[a4paper]{jpconf}
\usepackage{graphicx}
\usepackage{amsmath}
\usepackage{bbold}
\usepackage{epsfig}
\usepackage[latin1]{inputenc}

\def \begineq{\begin{equation}}
\def \endeq{\end{equation}}
\def \({\left(}
\def \){\right)}
\def \[{\left[}
\def \]{\right]}

\begin{document}
\title{The hard-core model on random graphs revisited}

\author{Jean Barbier$^\dag$, Florent Krzakala$^\dag$, Lenka Zdeborov\'a$^\star$ and Pan Zhang$^{\dag,\ddag}$}

\address{$\dag$ ESPCI and CNRS UMR 7083, 10 rue Vauquelin Paris 75005  France \\ Laboratoire de Physique Statistique, CNRS UMR 8550,  Universit\'e P. et M. Curie, Ecole Normale Sup\'erieure, Paris,  France}
\address{$\star$ Institut de Physique Th\'eorique, IPhT, CEA Saclay and URA 2306, CNRS, 91191 Gif-sur-Yvette, France.}
\address{$\ddag$ Santa Fe Institute, 1399 Hyde Park Road, Santa Fe, NM, USA}

\ead{jean.barbier@espci.fr, florent.krzakala@ens.fr, lenka.zdeborova@cea.fr, july.lzu@gmail.com}

\begin{abstract} We revisit the classical hard-core model, also known as independent set and dual to vertex cover problem, where one puts particles with a first-neighbor hard-core repulsion on the vertices of a random graph. Although the case of random graphs with small and very large average degrees respectively are quite well understood, they yield qualitatively different results and our aim here is to reconciliate these two cases. We revisit results that can be obtained using the (heuristic) cavity method and show that it provides a closed-form conjecture for the exact density of the densest packing on random regular graphs with degree $K\ge 20$, and that for $K>16$ the nature of the phase transition is the same as for large $K$. This also shows that the hard-code model is the simplest mean-field lattice model for structural glasses and jamming. 
 \end{abstract}

 Given a graph, how to put a large number of particles on the vertices avoiding any first-neighbor contact? This is the task one has to solve in the hard-core model (HC), also known as independent set (IS), or vertex cover (VC) when the meaning of covered/empty is switched. It is a NP-hard combinatorial optimization problem that has many applications such as scheduling problems \cite{Ambuhl}, inference of phylogenetic trees \cite{Auyeung04thelargest} or in the communications \cite{SafarCommunicationVC} where one seeks for the minimal set of placed sensor devices in a service area so that the entire area is accessible.

The hard-core model is defined as follows: We consider a graph $G=({V},{E})$  of size $N = |{V}|$ and associate an occupation number $\sigma_i \in \{0,1\}$ to every vertex $i \in {V}$, where $0$ stands for free and $1$ for occupied. The Gibbs measure that corresponds to the hard-core model reads 
\begineq
\label{eq.PS1}
P(\{\sigma_i\}_{i=1,\dots,N}) = \frac{1}{Z} e^{\mu \sum_{i \in {V} } \sigma_i}  \prod_{(ij) \in {E} } (1 - \sigma_i \sigma_j ) \, ,
\endeq
where $\mu$ is a constant called the chemical potential and $Z$ is a normalization called the partition function. The average packing fraction $\rho = 1/N \sum_{j=1}^N \sigma_j P(\{\sigma_i\}_{i=1,\dots,N})$ increases with $\mu$, the limit of $\mu\to \infty$ corresponding to the densest packing. 
The hard-core problem (\ref{eq.PS1}) is equivalent to the independent set problem where every edge is allowed at most one occupied end-node. This is dual (in the sense that $\sigma_i \to \neg \sigma_i$ and $\rho \to 1-\rho$) to the vertex cover problem, where every edge must have at least one occupied neighbor.

In statistical physics the hard-core model is seen as a very natural lattice model for the behavior of hard-spheres. The peculiar glassy properties of packed hard spheres are widely studied and their understanding has important physical consequences, for a review see e.g. \cite{parisi2010mean}.
The hard-core model is well studied in the statistical physics literature \cite{Weigt:481562,weigt2000num,HartmannWeigtStatMechaVC,hartmann01statisticalmechanics,hartmann2006phase,zhou2003vertex,journals/corr/abs-cond-mat-0605190}. 
 It was shown that as the density of covered sites increases the hard-core model on random graphs with small average degree undergoes a continuous phase transition (towards a so-called full replica symmetry breaking phase) \cite{RivoireThese,biroli2001lattice,rivoire:hal-00002292,mezard2007statistical,PhysRevE.80.021122}. 
For this reason it is often disregarded as a good model for the behavior of hard spheres, because one desires a discontinuous phase transition (towards the so called dynamical one step replica symmetry breaking phase) as the density of particles is increased in such a model. More complicated lattice glass models were hence introduced and studied \cite{biroli2001lattice,rivoire:hal-00002292}. In the class of models with a discontinuous phase transition there are in fact two such transitions, a dynamical one (also referred to as the mode coupling transition in the literature about glasses) and a condensation one (also called Kauzmann transition in the same literature) \cite{berthier2011theoretical}. 

In mathematics and computer science the hard-core model is also extensively studied. On large Erd\"os-R\'enyi or regular random graphs with constant average degree $K$ it has been shown that for sufficiently large $K$ there exists an independent set of density (fraction of covered nodes) $\rho \approx 2\log{K}/K$ \cite{bollobas1976cliques,frieze1990independence,dani2011independent}. While it is simple to design a polynomial algorithm that construct independent sets  (packings)  \cite{grimmett1975colouring} up to density $\rho = (1 + o(1))\log{K}/K$, about half the maximum size, no algorithm is known (either provably, or heuristically) to produce in polynomial time a packing of density $\rho = (1 + \epsilon)\log{K}/K$, with finite $\epsilon$. Hence a question of central importance in theory of algorithms is how to construct larger independent sets or what property of the problem makes it algorithmically so hard \cite{frieze1997algorithmic}. Recently the work of \cite{Coja-Oghlan:2011:ISR:2133036.2133048} shed light on this question by showing that for sufficiently large $K$ the space of all independent sets of density $\rho<\log{K}/K$ is simple (connected), for $\rho>\log{K}/K$ the space of independent sets splits into small cluster that are far away from each other. An exponential lower bound for a Metropolis process for sampling independent sets testifies that some onset of algorithmic difficulty happens at $\rho \approx \log{K}/K$ \cite{Coja-Oghlan:2011:ISR:2133036.2133048}. 
A remarkable recent work \cite{Gamarnik} showed, using the properties of clustering, that local algorithms cannot find dense configurations of the hard-core model. The existence of a discontinuous phase transition in the hard core model is also suggested by other  mathematical literature. Indeed, the results of \cite{Bhatnagar:2010:RTH:1886521.1886556} on the reconstruction threshold for the hard-core model indicates that the dynamical transition appears at densities $\approx \log{K}/K$, whereas the results of \cite{dani2011independent} indicate that both the condensation transition and the densest packing appear at $\approx 2 \log{K}/K$.

The main novel contribution of this paper is to settle down this apparent discrepancy between the physics and mathematical works concerning the nature of the phase transition in the hard-core problem on random graphs. The result is that whereas the current statistical physics picture with continuous phase transition is correct for graphs of small average degree ($K < 16$ for random regular graphs), the model has a discontinuous phase transition for degree $K>16$ and the large $K$ expansion of the corresponding equations is in agreement with all the mathematical results. This is showing that for large dimensional hard spheres (corresponding to large degree $K$) this simple hard-core model actually is a very good lattice model.

Another contribution of this paper is to point out that on the $K$-regular random graphs the one-step replica symmetry broken result for the size of largest independent set, or minimal vertex cover, is stable towards more steps of replica symmetry breaking for $K\ge 20$, and hence for $K\ge 20$ we provide a closed form conjecture about the {\it exact} values of this density. 

We structure the paper as an overview of the results obtained for the hard-core model after a careful investigation of the cavity solution of the model on random regular graphs of degree~$K$. We used the heuristic cavity method that is well known in statistical physics and has been derived and explained in details elsewhere, see e.g. \cite{Zdeborova09} for an overview. In order to keep this paper accesible also to readers not familiar with the cavity method we give only the resulting expressions and mostly only in cases when they are simple to state and lead to interesting conjectures about the structure of dense packings or the density of the largest packing.

\section{The densest packing, largest independent set and minimal vertex cover}

\subsection{Replica symmetric result}
We define the cavity message $\pi_{i \to j}$ to be the probability that a node $i$ is occupied in the cavity graph where the edge $(ij)$ has been deleted. In this modified graph, we then assume that other neighbors of node $i$ are independent and write the belief propagation equations 
\begin{equation}
\label{eq.RS2}
\pi_{i \to j}=  \frac{e^{\mu}\prod_{k \in \partial i \backslash j} (1-\pi_{k \to i})}{1 + e^{\mu}\prod_{k \in \partial i \backslash j} (1-\pi_{k \to i})} \, ,
\end{equation}
where $\partial i \backslash j$ are all the neighbors of node $i$ excluding $j$. 
On a graph $\mathcal{G}$ the Bethe free energy is given using the fixed point of eq. (\ref{eq.RS2}) as
\begin{equation}
\label{eq.RS4}
f_{RS}^{\mathcal{G}}  =
-\sum_{i \in {V} }\log{[1 + e^{\mu}\prod_{j \in \partial i}(1- \pi_{j \to i} ) ] } +\sum_{(ij) \in E}\log{(1 - \pi_{j \to i}\pi_{i \to j} ) }\, .
\end{equation}

On random $K$-regular graphs the above equation is satisfied by $\pi_{i \to j} = \bar\pi $ that solves the following equation
\begin{equation}
\label{eq.RS3}
\bar\pi = \frac{e^{\mu} (1-\bar\pi)^{{K-1}}}{1 + e^{\mu}(1-\bar\pi)^{{K-1}} }=e^{\mu} (1-\bar\pi)^K\, .
\end{equation}
The corresponding densities of free energy, density, and entropy reads (at the fixed point):
\begin{align}
\label{eq.RS5}
&f_{RS}(\mu) = -\log{(1 + \bar\pi) } + \frac{K}{2} \log{(1-\bar\pi^2)} \, ,\\
\label{eq.RS6}
&\rho_{RS}(\mu) = - \frac{\partial f^{\mathcal{G}}_{RS}}{\partial \mu} \Big|_{\bar{\pi}} = \frac{\bar\pi}{1+ \bar\pi} \, ,\\
\label{eq.RS7}
&s_{RS}(\rho) = -f_{RS} -\mu \rho_{RS}\, .
\end{align}
where again $\rho$ is the particle density. Solving $s_{RS}(\rho^{\rm RS}_s)=0$, we get the replica symmetric (RS) result for the densest packing evaluated in Table~\ref{Tab1}. Due to a general property that in the absence of quenched disorder the replica symmetric solution on random regular graphs is equivalent to the first moment (annealed) calculations \cite{MoraThese} this replica symmetric result agrees with the rigorously known upper bounds. 

However, as in many other optimization problems the replica symmetric result is not exact. A necessary condition for its correctness is referred to as the local stability. The condition that needs to be satisfied for the RS approach to be stable is \cite{RivoireThese,PhysRevE.80.021122} 
\begin{equation}
\label{eq.RS9}
\lambda = (K-1)  \left( \frac{\partial \pi_{i \to j}} {\partial \pi_{k \to i}} \right)_{\bar \pi}^2 
=  (K-1) \bar{\pi}^2<1\, .
\end{equation}
Using (\ref{eq.RS7}) we get the local stability threshold of the RS solution on random $K$-regular graphs
\begin{equation}
\label{eq.RS10}
\mu_l = -\frac{1}{2}\log({K-1}) - K \log(1-\frac{1}{\sqrt{{K-1}}}), \quad  
\rho_l = \frac{1}{1 +\sqrt{{K-1}}} \, .  
\end{equation}
We hence conclude that close to the densest packings, the replica symmetric results for the densest packing are never stable and hence not exact. They only provide an upper bound. We will explain further the physical meaning of the density $\rho_l$ in Section \ref{KS}. 

\begin{table}[!ht]
\begin{center}
\begin{tabular}{|c|c|c|c|c|}
\hline
K  &  $\rho^{\rm RS}_l$ & $\mu^{\rm RS}_l$ & $\rho^{\rm RS}_s$ & $\rho^{\rm 1RSB}_s$\\
\hline
\hline
3 & 0.4142 & 3.3373 & 0.4591 & 0.4509\\
4 & 0.3660 & 2.8955  & 0.4206 & 0.4112 \\
5 & 0.3333 & 2.7726 & 0.3887 & 0.3793\\
6 & 0.3090 & 2.7520 & 0.3620 & 0.3530\\
7 & 0.2899 & 2.7768 & 0.3394 & 0.3309\\
8 & 0.2743 & 2.8251 & 0.3200 & 0.3120\\
9 & 0.2612 & 2.8867 & 0.3031 & 0.2956\\
10 & 0.2500 & 2.9560 & 0.2882 & 0.2811\\
15 & 0.2109 & 3.3450 & 0.2338 & 0.2286\\
\hline
\end{tabular}
\caption{\label{Tab1} The replica symmetric clustering transition given by (\ref{eq.RS10}), the replica symmetric densest packing $\rho^{\rm RS}_s$ and the 1RSB densest packing $\rho^{\rm 1RSB}_s$ which are obtained from (\ref{eq.RS7}) and (\ref{eq.SP3}) respectively.}
\end{center}
\end{table}

\subsection{1RSB: exact result for $K\ge 20$}
\label{exact}

The technique that provides next candidate for the densest packing is derived from the survey propagation (SP) equations  (also called energetic 1RSB cavity equations) \cite{mezard:hal-00002362}. For vertex cover (equivalent to the hard-core model up to a dual transformation) on random graph the survey propagation equations were first derived in \cite{RivoireThese,hartmann2006phase}, and read
\begin{equation}
q_{i \to j} \(\{{  q_{k \to i}  \}}\) = \frac{e^{y} \prod_{k\in \partial i \setminus j} (1-q_{k \to i})}{1+(e^{y}-1) \prod_{k\in \partial i \setminus j} (1-q_{k \to i})}\, ,
\label{SP_equation}
\end{equation}
where $q_{i \to j}$ is the (cavity) fraction of vertices that are frozen to a value $1$ (or ``occupied'') within clusters. For random $K$-regular graphs the survey propagation equations admit a factorized solution with a fixed point
\begin{equation}
\label{eq.SP2}
\bar{q} = \frac{e^{y}(1-\bar{q})^{{K-1}}}{1+(e^{y}-1)(1-\bar{q})^{{K-1}}}\, .
\end{equation}
This leads to the following complexity (i.e. logarithm of number of clusters of solution):
 \begin{eqnarray}
\label{eq.SP3}
\Sigma(y) &= \frac{y e^{-y}}{Z^{f}} \left[1-(1-\bar{q})^{K} \right]  - \frac{Kye^{-y}\bar{q}^2}{2Z^{e}}+\log(Z^{f})-\frac{K}{2} \log(Z^{e}) \, ,\\
\text{with} & \ Z^{f} = (1 - \bar{q})^K + e^{-y}[1 - (1 - \bar{q})^K], \ Z^{e} = 1 + \bar{q}^2(e^{-y}-1)\, .
 \end{eqnarray}
The 1RSB result for the densest packing (largest independent set) is then given as 
\begin{equation}
\label{eq.SP5}
\rho^{\rm 1RSB}_s = \frac{e^{-y_0}[1-(1-\bar q)^K]}{Z^f} -\frac{Ke^{-y_0} \bar{q}^2}{2Z^{e}}\, ,
\end{equation}
where $y_0$ is such that the complexity $\Sigma(y_0) = 0$. The values obtained by evaluating these expressions for $K\le 100$ are summarized in Tables~ \ref{Tab1} and \ref{Tab2}. In general one needs to take into account more levels of replica symmetry breaking in order to obtain exact results for the densest packing. The 2RSB calculation is, however, somewhat involved and was done in \cite{RivoireThese} in the $K=3$ case, in which case one obtains $\rho^{\rm RS}_s = 0.45906$, $\rho^{\rm 1RSB}_s =0.45086$, $\rho^{\rm 2RSB}_s =0.45076(7)$.

Note that the Franz-Leone-Panchenko-Talagrand theorem implies that this 1RSB result is an upper bound \cite{FranzLeone03b,panchenko2004bounds}. We shall now argue when the 1RSB solution is stable against perturbation towards a 2RSB solution for large enought $K$. While this does not prove the absence of a higher RSB solution, it is a very good hint that the 1RSB solution is the exact one, and we thus conjecture that it provides {\it exact} values for the densest packing for $K\ge 20$.
\begin{table}[!ht]
\begin{center}
\begin{tabular}{|c|c|c|c|c|c|c|c|c|}
\hline
K & $\rho_{d}$ & $\rho_{c}$ & $\rho_l$ & $\rho^{\rm RS}_s$ & $\rho^{\rm 1RSB}_s$ & $\mu_{d}$ & $\mu_{c}$ & $\mu_l$\\
\hline
\hline
16 & 0.2051 & 0.2051 & 0.2052 & 0.2257 & 0.2207 & 3.421 & 3.421 & 3.425 \\
17 & 0.1989 & 0.1989 & 0.2000 & 0.2182 &   0.2135 & 3.459 & 3.459 & 3.504 \\
18 & 0.1933 & 0.1933 & 0.1952 & 0.2113&   0.2068 & 3.502& 3.502 & 3.583\\
19 & 0.1880 &  0.1880& 0.1907 & 0.2048 &   0.2006 & 3.539 &3.539  & 3.662\\
\hline
20 & 0.1830 & 0.1833 & 0.1866 &0.1989 & 0.1948 & 3.576 & 3.591 & 3.740 \\
30 & 0.1453 & 0.1465 & 0.1566 &0.1555 & 0.1529 & 3.818 & 3.886 & 4.479 \\
40 & 0.1214 & 0.1231 & 0.1380 &0.1292 & 0.1273 & 3.966 & 4.086 & 5.148 \\
50 & 0.1047 & 0.1068 & 0.1250 &0.1113 & 0.1098 & 4.072 & 4.241 &5.762 \\
60 & 0.0924 &  0.0947 & 0.1152 & 0.0981& 0.0970& 4.155 & 4.369 & 6.330 \\
70 & 0.0828 & 0.0853 & 0.1075 &0.0881  & 0.0871 & 4.221 & 4.479 & 6.862 \\
80 & 0.0752 & 0.0778 & 0.1011 &0.0800  & 0.0792 & 4.279 & 4.575  & 7.364 \\
90 & 0.0690 & 0.0716 & 0.0958 &0.0735  & 0.0728 &  4.329 & 4.658  & 7.840 \\
100 & 0.0638 & 0.0664 & 0.0913 &0.0680& 0.0674 &  4.372 & 4.736 & 8.294 \\
\hline
\end{tabular}
\caption{\label{Tab2} Values for the dynamical $\rho_d$, condensation $\rho_c$, RS stability $\rho_l$ transitions (and the corresponding values of $\mu$), and for the RS and the 1RSB densest packings, the last is exact for $K\ge 20$. There is an estimated numerical error $\pm 1$ on the last digit for the values of $\rho_d$, $\mu_d$, $\rho_c$, $\mu_c$ since those were evaluated by by population dynamics with population size $N=10^5$, equilibration time $T_{eq}=10^3\times N$  and averaging time $T=10^4\times N$. Given this numerical solution, we are not entirely certain that the transition at $K=16$ is not actually still a second order one. Data for $K\ge 20$ are separated by a line because they are in the regime where the 1RSB solution is expected to be exact (see section~\ref{exact}).}
\end{center} 
\end{table}
These two stability checks have been introduced and used extensively by \cite{muller2004glassy,rivoire:hal-00002292,RivoireThese,mertens2006threshold,krzkakala2004threshold,Zdeborova09} and we refer to these work for more details. In the present case, they have been considered in \cite{RivoireThese,PhysRevE.80.021122} but evaluated only for small values of $K$.

\subsubsection{SP type-I stability}
The first type of instability to check is if the fixed point of eq.~(\ref{SP_equation}) is stable against small perturbations on a given single instance. This is the case as long as 
\begin{equation}
(K-1) \[ \frac{\partial  q_{i \to j} \(\{{  q_{k \to i}  \}}\)   }{\partial q_{k_1 \to i}} \Big|_{\bar{q}}  \]^2 = (K-1) \[ \frac {\bar q}{1-(1- \bar q)^{d-1}}\]^2<1\, .
\end{equation}
Performing the numerical computation at the value of $y_0$ such that the complexity vanishes, we find that the SP solution is stable as long as $K>3$ (and is unstable for $K=3$). 

\subsubsection{SP type-II stability} A more involved stability computation is the Type-II stability, also called bug proliferation, that has been considered first by Rivoire for the present problem \cite{RivoireThese}. Suppose that a message of type $\sigma=0/1$ (empty/occupied) is turned into another message $\tilde\sigma=1/0$ with small probability $\pi^{\sigma\to\tilde\sigma} \ll 1$. Then, some output messages may change, and the system is said unstable with respect to Type-II perturbations if such a bug propagates through the whole system. In the linear response regime, for a generic change of the first input message from the value $\tau$ to $\tilde{\tau}$ in eq.~(\ref{SP_equation}) happening with probability   $\pi^{\sigma\to\tilde\sigma}_1$  we can calculate the probability of changing the output message   $\pi^{\tau\to\tilde\tau}_0 $ from the value $\tau$ to $\tilde\tau$ as 
\begin{equation}
  \label{eq:type2}
  \pi^{\tau\to\tilde\tau}_0 = \frac 1{\cal C} 
  \sum_{\substack { (\sigma,\sigma_2,...,\sigma_d) \to \tau \\
   (\tilde\sigma,\sigma_2,...,\sigma_d) \to \tilde\tau } } 
  \pi^{\sigma\to\tilde\sigma}_1 
  \eta^{\sigma_2}_2 \cdots  \eta^{\sigma_d}_d \ \ 
  e^{y \bar{\tau}} \, ,
\end{equation}
where ${\cal C} $ is the denominator in eq.~(\ref{SP_equation}), $e^{y \bar{\tau}}$ is the "reweighting" connected to an energy change that appears in the survey propagation equations, and the $\eta$ are edge-variables related with the SP ones such that $\eta^1=q$ and $\eta^0=1-q$.  This defines a matrix $V$ with entries
\begin{equation}
\label{eq:V}
  V_{\tau\to\tilde\tau,\sigma\to\tilde\sigma} = \frac
  {\partial\pi^{\tau\to\tilde\tau}_0 }{\partial
    \pi^{\sigma\to\tilde\sigma}_1 } \ ,
\end{equation}
evaluated at the 1RSB solution. Noting $\lambda$ the largest eigenvalues of this matrix, the system is then stable if $(K-1)\lambda<1$. In the present case, the matrix reads
\begin{equation}
\label{eq.SP6}
\begin{bmatrix}
V_{0 \to1, 0 \to1} & V_{0\to1,1\to0} \\
V_{1 \to0, 0 \to1} & V_{1\to0,1\to0} \\
\end{bmatrix}
=
\begin{bmatrix}
0 & \bar{q}/(1-\bar q)\\
e^{-y/2} \bar{q}/(1-\bar q) & 0
\end{bmatrix} \, .
\end{equation}
SP is thus type-II stable as long as \cite{RivoireThese}:
\begin{equation} 
(K-1)^2\(\frac{\bar{q}}{1-\bar{q}}\)^2e^{-y}<1\, ,
\end{equation}
which, for the values of $y_0$ corresponding to the packing transition, is verified for $K\ge20$.

From these two checks, we thus conclude that the 1RSB value is consistent as long as $K\ge20$, and conjecture that it gives the exact values for densest packings on random regular graphs. In fact, from what has been observed in other models (see for instance \cite{krzakala2008potts} for the coloring problem) we expect the 1RSB approach to be exact for the whole range of the chemical potential $\mu$ in this case. In fact, this seems to be consistent with the numerical results of \cite{PhysRevE.80.021122}. Proving this conjecture  for finite $K$ remains a challenging problem.

\section{The clustering and condensation phase transitions}

The structure and properties of dense packings is interesting even for densities well below the largest one, as proven for $K$ large enough in \cite{Coja-Oghlan:2011:ISR:2133036.2133048,Bhatnagar:2010:RTH:1886521.1886556}. Statistical physics of models such as the present one describes two important phase transitions - clustering and condensation \cite{krzakala2007gibbs}. In the physics of glassy materials the clustering transition corresponds to the so-called dynamical or mode coupling transition and the condensation transition to the Kauzmann transition. In some models these two transitions coincide, and many authors claimed that the hard-core model is in this class \cite{RivoireThese,biroli2001lattice,rivoire:hal-00002292,mezard2007statistical,PhysRevE.80.021122}. However, the rigorous results of \cite{Coja-Oghlan:2011:ISR:2133036.2133048,Bhatnagar:2010:RTH:1886521.1886556} clearly suggest that for large $K$ the clustering and condensation phase transitions are distincts and far away from each other ($\log{K}/K$ versus $2\log{K}/K$).

\subsection{Continuous clustering transition}
\label{KS}

When the clustering and condensation transitions coincide then they also coincide with the local stability of the replica symmetric solution eq.~(\ref{eq.RS10}). In such a case the density $\rho_l$ marks a point below which the replica symmetric solution is exact. In particular below which the annealed (first moment) calculation counts correctly the number of configurations ${\cal N}(\rho)$ at a given density $\rho$ in the sense that 
\begin{equation} 
     s_{\rm RS}(\rho) =  \lim_{N\to \infty} \log{\langle {\cal N}(\rho) \rangle } /N  =   \lim_{N\to \infty} \langle \log{{\cal N}(\rho)} \rangle /N = s(\rho) \, .
\end{equation}
Above $\rho_l$ the number of configurations of a given density is strictly smaller that the replica symmetric result. There is a non-analyticity in the function $s(\rho)$ at $\rho_l$. 
Note that the local stability of the replica symmetric solution is equivalent to the Kesten-Stigum bound for the reconstruction on graphs \cite{KestenStigum66b}.

\subsection{Discontinuous clustering transition and condensation transition}

In order to investigate whether the clustering transition is distinct from the condensation one and from the $\rho_l$ transition, we need to investigate the 1RSB solution at the value of Parisi parameter $m=1$. This is again a rather involved heuristic method, that is well known and established in statistical physics \cite{IPC.Mezard.Montanari}. For vertex cover the corresponding equations were studied in \cite{PhysRevE.80.021122}. The 1RSB approach at $m=1$ on random regular graphs in the absence of quenched disorder simplifies significantly due to its relation to the properties of the planted ensemble \cite{krzakala2009hiding}.

To locate the dynamical transition $\rho_d$ we need to do the following: We plant a configuration of density corresponding to the replica symmetric result. We initialize belief propagation in this configuration and iterate till convergence. The density above which the resulting BP fixed point starts to be correlated with the initial condition marks the clustering transition. To measure the level of correlation we define a so-called overlap between the fixed point and the initial configuration. When $\rho_d \neq \rho_l$ the overlap changes discontinuously and we speak about a 1st order phase transition, whereas when $\rho_d=\rho_l$ it changes continuously and we speak about a second order phase transition. The density above which the Bethe free energy of the resulting BP fixed point becomes smaller than the replica symmetric free energy marks the condensation transition~$\rho_c$. Indeed, the difference between these two values is simply the "complexity'' of the equilibrium states, that is, the logarithm of the number of equilibrium states $\Sigma$ (per spins). When $\Sigma(s)=0$, the condensation transition appears and a non-extensive number of clusters of solutions starts to dominate the Gibbs measure. The local stability of the replica symmetric solution at $\rho_l$ marks the density above which it become algorithmically easy to find the planted configuration.

In \cite{PhysRevE.80.021122} the authors realized that for $K < 16$ the three transitions coincide, $\rho_d=\rho_c=\rho_l$, whereas for $K >16$ one has $\rho_l > \rho_d$. They, however, did not observe $\rho_d \neq \rho_c$. Note that if $\rho_l > \rho_d$ then the transition must be of 1st order with $\rho_c > \rho_d$ even if the difference might be numerically indistinguishable. Indeed, if the transition was continuous then we could expand around the replica symmetric solution to describe the behavior around the transition, but such an expansion is exactly what leads to the computation of $\rho_l$, hence either we have $\rho_d=\rho_c=\rho_l$ or $\rho_d< \rho_c < \rho_l$. 

In the present work we hence evaluated the values of $\rho_d$ and $\rho_c$ more carefully and found that indeed for $K > 16$ the transition is discontinuous (and our data indicates that it is likely to be the case as well for $K=16$ although this is difficult to ascertain), and for $K \ge 20$ we were able to see numerically a clear difference between $\rho_d$ and $\rho_c$, the values are summarized in Table \ref{Tab2}.

From a large $K$ expansion of the corresponding equations we found that $\rho_c \approx 2 (\log{K}  -  \log{\log{K}} +1-\log 2)/K + o(1/K)$ which agrees in these 3 orders with the expansion of the 1RSB result for the densest packing. This is no surprise, because these 3 orders can be obtained even from the rigorously known lower and upper bound of \cite{dani2011independent} and it is well known that the second moment technique that  \cite{dani2011independent} used cannot surpass the condensation transition. For the large $K$ scaling of the dynamical transition we obtained $\rho_d \approx \log{K}/K + o(\log{K}/K)$ which also agrees with the rigorous results in the clustering \cite{Coja-Oghlan:2011:ISR:2133036.2133048} and with the results of \cite{Bhatnagar:2010:RTH:1886521.1886556} on reconstruction on trees.

\begin{figure}[!ht]
\hspace{-0.2cm}
\includegraphics[width=0.52\textwidth]{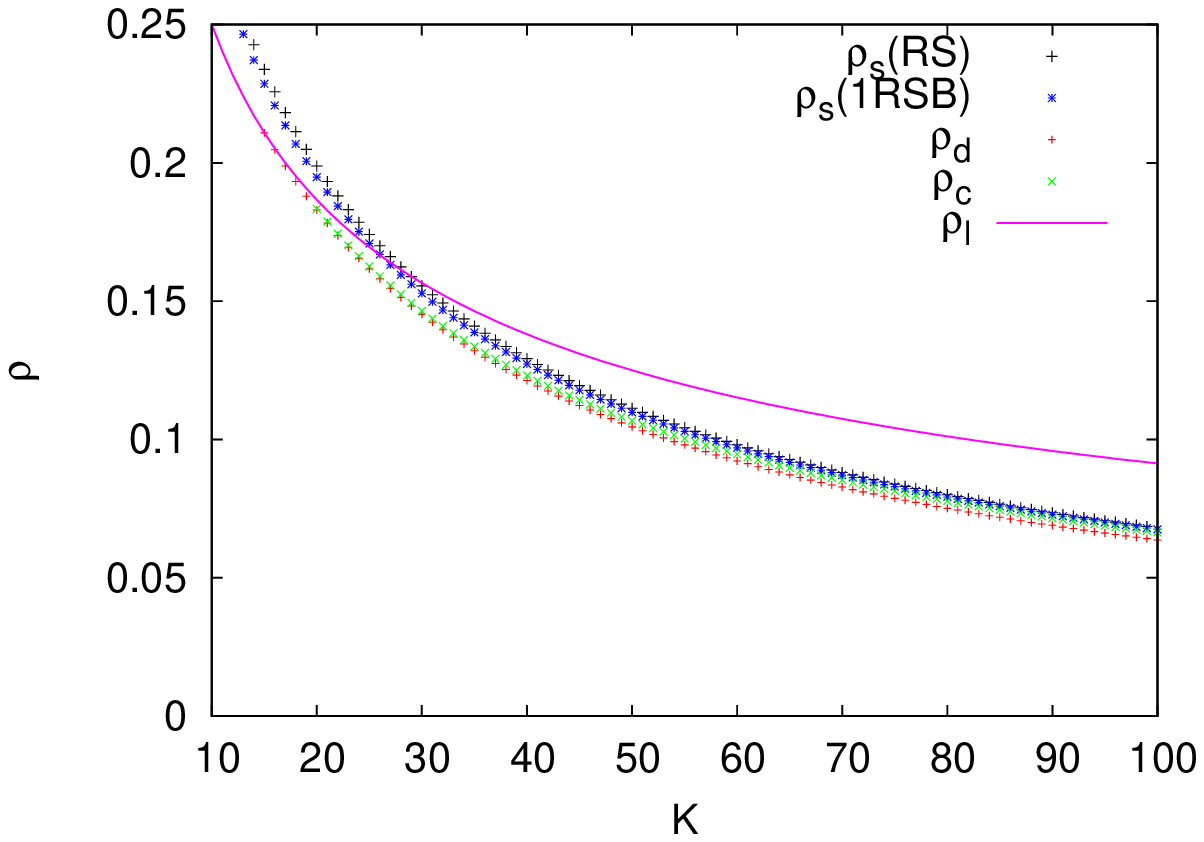}
\hspace{-0.5cm}
\includegraphics[width=0.52\textwidth]{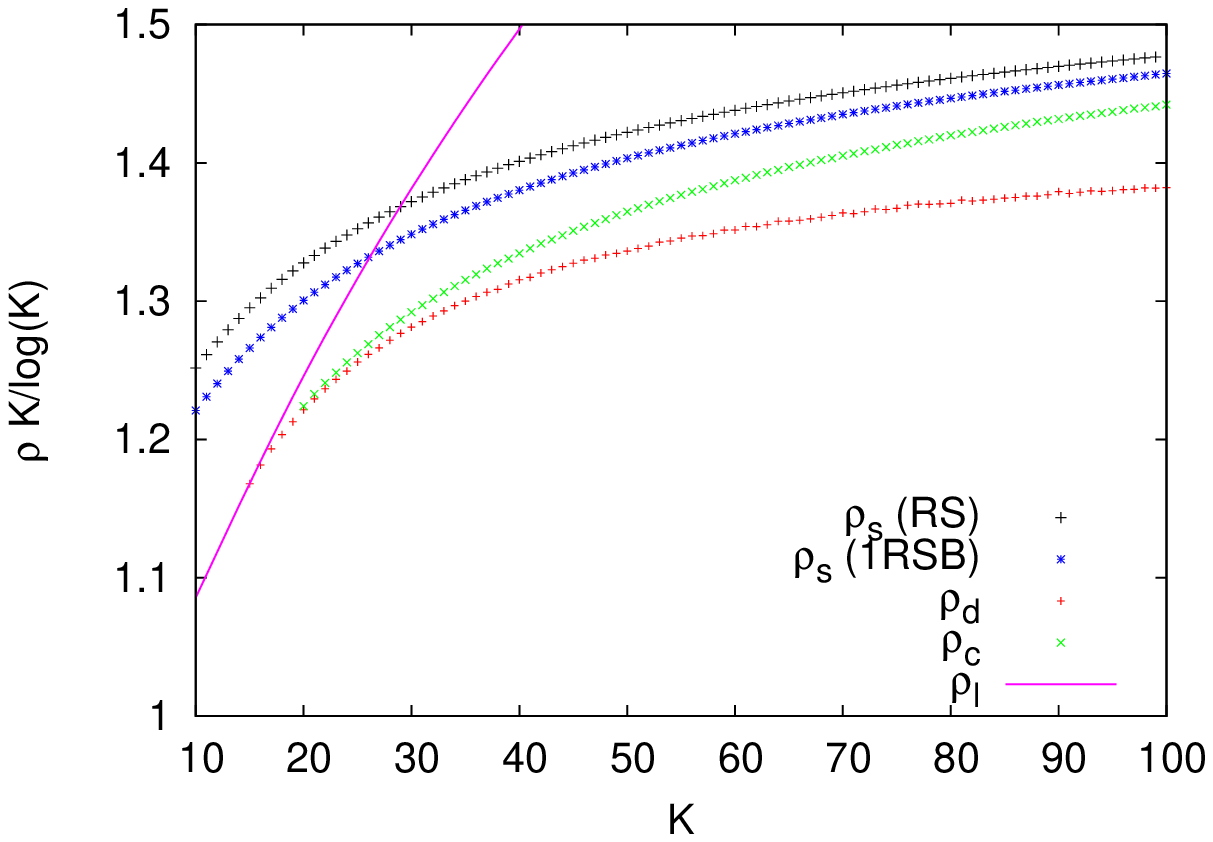}
\caption{\label{label} Phase diagram of the hard-core model on $K$-regular random graphs. The critical densities for clustering $\rho_d$, condensation $\rho_c$, replica symmetric stability $\rho_l$, replica symmetric densest packing $\rho_s^{\rm RS}$ and 1RSB densest packing $\rho_s^{\rm 1RSB}$. The main result of this paper is that the dynamical and condensation transitions are distinct for $K > 16$ and the 1RSB results are exact for $K\ge 20$. The right part of the figure is the rescaled density $\rho K /\log{K} $ plotted as a function of $K$. We can see that even for $K=100$ we are still extremely far from the asymptotic values $\lim_{K\to \infty} \rho_d K /\log{K} \to 1$, and $\lim_{K\to \infty} \rho_c K /\log{K} \to 2$.}
\end{figure}


\section{Conclusion}
\label{conclusion}
In this paper, we matched the two extreme regimes of the hard-core model on random regular graphs to give a picture of the behavior of the solution space for the full range of average degrees. For small degrees, the RS solution is valid at low enough chemical potential and becomes continuously unstable towards a full-RSB solution as $\mu$ increases. When $K\ge16$, however, the picture changes: A stable 1RSB solution appears above a sharp dynamical transition and then, as $\mu$ continues to increase, a condensation over a finite set of clusters of the probability measure happens. A full-RSB solution  is still found for large enough $\mu$ until, for $K\ge20$, even the densest packing is 1RSB stable. We expect this picture to be qualitatively similar for random graphs with other degree distributions (with finite second moment), although the numbers will change slightly. We hope our paper will help to clarify the different pictures and results in the present literature, and stimulate rigorous work towards proving these conjectures.

Another interesting remark is that since the phase transition is, for large enough $K$, of the random first order type type, it has the glassy phenomenology associated with hard spheres in physics \cite{parisi2010mean}. In fact, by associating the graph degree with the kissing number via a very rough estimate $K \approx 2^d$, 
the hard-core dynamical transition arises at $\rho_d \propto d/2^d$ which coincide with the scaling for the dynamical transition in hard spheres in large dimension \cite{parisi2010mean}. The analogy extends less nicely to the scaling of the condensation and densest packings where for hard spheres one expects $\rho \propto d\log{d}/2^d$ \cite{parisi2010mean}, but these are entropy driven quantities and hence differences between continuous and discrete models are to be expected (there are no small vibration in the discrete description). Still this rough agreement is remarkable and the hard-core model should therefore be seen as the simplest mean field lattice model for the glass and jamming transitions.

\section*{Acknowledgments}
This work has been supported in part by the ERC under the European Union's 7th Framework Programme Grant Agreement 307087-SPARCS, by the Grant DySpaN of "Triangle de la Physique" and by the French Minist\`ere de la defense/DGA.

\newpage

\section*{References}

\bibliographystyle{iopart-num.bst}
\bibliography{refs}

\end{document}